\documentclass[aps,prl,twocolumn,groupedaddress,showpacs]{revtex4}

\usepackage{graphicx}
\usepackage{dcolumn}

\begin{document}


\title{Lifetime Widths of Surface States on Magnetic Lanthanide Metals}

\author{A.\ Bauer, D.\ Wegner, and G.\ Kaindl}

\affiliation{Institut f{\"u}r Experimentalphysik, Freie
Universit{\"a}t Berlin \\  Arnimallee 14, 14195 Berlin-Dahlem,
Germany}

\date{February 23, 2005}

\begin{abstract}
Low-temperature scanning tunneling spectroscopy is used to study
electronic structure and dynamics of $d$-like surface states of
trivalent lanthanide metals from La to Lu. The magnetic exchange
splitting of these states is found to scale with the 4\emph{f}
spin multiplied by an effective exchange-coupling constant that
increases with 4\emph{f} occupancy in an approximately linear way.
The dynamics of the surfaces states, as revealed by the lifetime
width, is dominated by electron-phonon scattering in the occupied
region and by electron-magnon scattering in the unoccupied region,
respectively.
\end{abstract}


\maketitle

%
%
Electronic structure and dynamics of surface states have recently
attracted substantial attention. For the noble metals, a profound
understanding of binding energies and lifetimes of surface states
has been achieved, mainly due to improvements in the experimental
and theoretical techniques \cite{Ech04}. Here, for defect-free
surfaces the lifetime of a surface state is predominantly
determined by electron-phonon (e-ph) and electron-electron (e-e)
scattering rates, with the latter including both intraband and
interband excitations \cite{Vit03}. On the other hand, much less
is known on surface-state dynamics in transition metals,
particularly in the lanthanides. Across the series, the lanthanide
metals have similar conduction-band structures but different
4\emph{f} occupancies, $n_{4f}$, and hence different
4\emph{f}-spin moments that interact with the ($5d6s$)-conduction
electrons and the surface state by exchange interaction. Below
specific temperatures, all lanthanide metals with non-vanishing
4\emph{f} moment exhibit long-range magnetic order that causes an
exchange splitting of the electronic states
\cite{Kur02,Bod98,Bau02,Sch00c}. For these magnetic transition
metals, collective excitations of spin waves (magnons) provide
additional decay channels for excited electronic states
\cite{Zhu04}. Electron-magnon (e-m) scattering processes, where an
electron undergoes a spin-flip and a magnon is either created or
annihilated, are difficult to distinguish from e-ph scattering
processes since both are expected to exhibit similar temperature
dependences \cite{Skr90}.

The electronic structure of surface states and their dynamics can
be studied by various techniques including photoemission (PE),
inverse photoemission (IPE), two-photon-photoemission (2PPE), and
scanning tunneling spectroscopy (STS) \cite{Ech04}. While 2PPE is
a time-resolved technique that can directly determine relaxation
times, PE, IPE, and STS measure spectral lifetime widths of
electronic states, i.e.\ inverse lifetimes.

In this Letter, we report on a systematic low-temperature STS
study of the $5d_{z^{2}}$-like surface states on the (0001) faces
of trivalent lanthanide metals. The magnetic exchange splitting,
$\Delta_{ex}$, of these laterally localized surface states is
found to scale with the 4\emph{f} spin moment multiplied by an
effective exchange-coupling constant that increases with $n_{4f}$
in an approximately linear way due to lanthanide contraction.
Concerning the dynamics of the surface states, we find much
shorter lifetimes for the (unoccupied) minority states as compared
to the (occupied) majority states. While the lifetimes of the
occupied states are mainly determined by e-ph scattering, the
lifetimes of the unoccupied states are dominated by e-m
scattering.

%
%

The experiments were performed in an ultrahigh vacuum (UHV)
chamber equipped with a low-temperature scanning tunneling
microscope (STM) operated at 10 K \cite{Bau02}. The samples were
prepared {\em in-situ} by electron-beam evaporation of the
lanthanide metals La ($4f^{0}$), Nd ($4f^{3}$), Gd ($4f^{7}$), Tb
($4f^{8}$), Dy ($4f^{9}$), Ho ($4f^{10}$), Er ($4f^{11}$), Tm
($4f^{12}$), or Lu ($4f^{14}$), respectively, and deposition on a
clean W(110) single crystal kept at room temperature.
Subsequently, the samples were annealed at temperatures between
 500 and 1000 K, resulting in smooth hcp (0001) films (dhcp
in case of La and Nd) with atomically flat terraces, and
local-thickness variations of only a few monolayers (ML)
\cite{Bau02}. The average thickness of the deposited film was
monitored with a quartz microbalance. STS spectra were recorded
with fixed tip position and switched-off feedback control. The
differential conductivity, $dI/dU$, with $I$ being the tunneling
current and $U$ the sample bias voltage, was measured as a
function of $U$ by modulating $U$ and recording the induced
modulation of $I$ via lock-in technique. $dI/dU$ is proportional
to the density of states at the sample surface in good
approximation. A modulation amplitude of 1 mV (rms) at a frequency
of $\approx 360\, \mbox{Hz}$ was used, with the time constant of
the lock-in amplifier set to 100 ms, at a sweep rate of $\approx
6\, \mbox{mV/s}$. To correct for binding energy shifts due to the
finite time constant, the spectra were recorded in both
directions, from lower to higher and from higher to lower sample
bias. Since both the STM tip and the sample were cooled to 10 K,
the energy resolution was $\approx 3\, \mbox{meV}$, corresponding
to $3.5\, k_{B}T$.

%
%

\begin{figure}
\begin{center}
\includegraphics{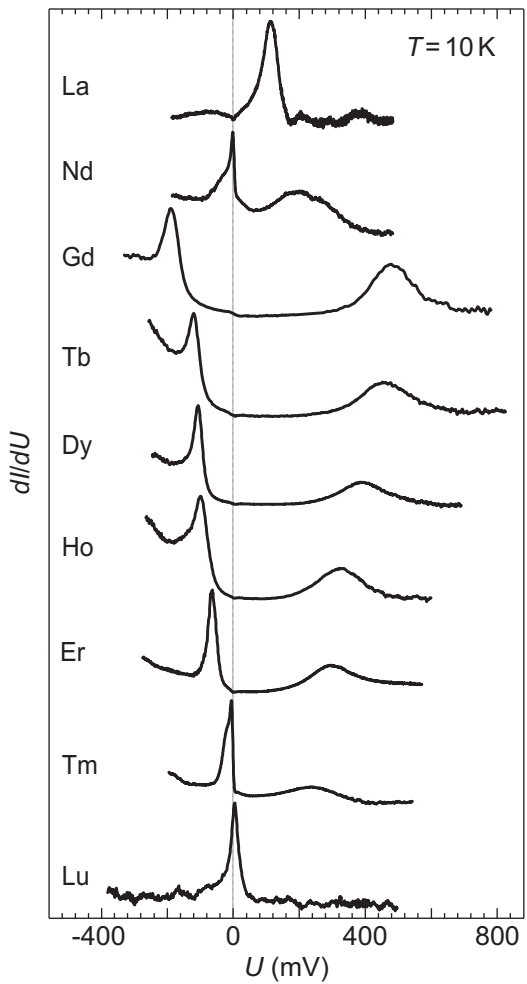}
\caption{\label{fig1} STS spectra of (0001)-surface states of the
trivalent lanthanide metals recorded at 10~K. The spectra are
overview spectra. Better resolved spectra of each peak were used
for data analysis, as described in detail in Ref.\ \cite{Bau02}.}
\end{center}
\end{figure}

Fig.~\ref{fig1} shows tunneling spectra of the trivalent
lanthanide metals La, Nd, Gd, Tb, Dy, Ho, Er, Tm, and Lu. The
\emph{d}-like surface states are clearly resolved as pronounced
peaks. For the lanthanides with non-vanishing 4\emph{f}-spin
moments, majority and minority exchange-split spin states appear
below and above $E_{F}$ ($U = 0$), respectively. For La and Lu,
with $n_{4f}=0$ and $n_{4f}=14$, respectively, the spectra exhibit
a single peak. The strongly peaked density of surface states
probed by STS is the consequence of a high degree of lateral
localization, resulting in small dispersions of the surface-state
bands with effective masses $m^{*}/m \gg 1$ \cite{Bau02,Sch03}.
For Nd and Tm, with the majority surface states rather close to
$E_{F}$, and possibly also for Lu, additional sharp features
directly at $E_{F}$ show up in the STS spectra that represent
many-body resonances due to an interaction of the localized
surface state with bulk conduction electrons; hence, these
features can be described by Fano line shapes \cite{Fan61,Weg04}.

Due to the small dispersions of the surface-state bands, the
surface-state peaks in the tunneling spectra can be described by
Lorentzian lines, with widths $\Gamma$ that are related to
lifetimes $\tau$ by $\Gamma = \hbar / \tau$. Only for narrow
peaks, small deviations from a symmetric line shape due to
dispersion are recognizable. Since the surface-state bands show a
downward dispersion with a band maximum at $\overline{\Gamma}$
(with the exception of Lu, for which an M-shaped dispersion had
been calculated \cite{Weg04}), the trailing edges of the peaks in
the STS spectra are generally steeper than the leading edges
\cite{Bau02}. In order to account for this asymmetry, the peaks
were fitted with adequate superpositions of Lorentzian lines; a
detailed description of the fit procedure is given in
Ref.~\cite{Bau02}. Fit results for energies and linewidths in the
present work refer to surface states at the band maximum. In case
of Nd and Tm, the trailing edges of the surface-state peaks are
superimposed on the mentioned resonances at $E_{F}$, and the
asymmetries of the peaks cannot be determined unambiguously.
Therefore, single Lorentzian lines were fitted to these peaks. The
errors that result from a neglect of surface-state dispersion are
less than 10 meV for both the energies and the widths of the
surface state \cite{Bau02}.

\begin{figure}[b]
\begin{center}
\includegraphics{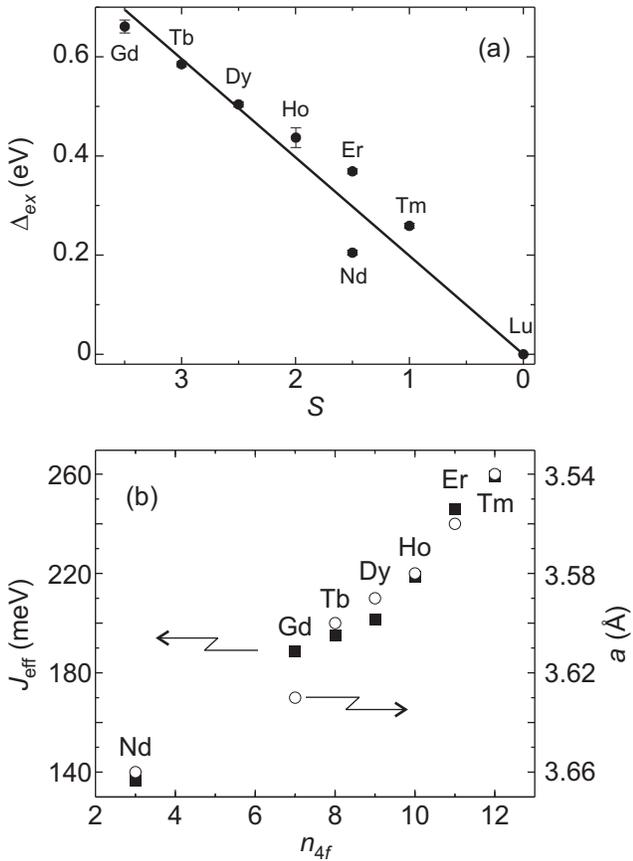}
\caption{\label{fig2} (a) Exchange splitting of (0001)-surface
states, $\Delta_{ex}$, at 10 K plotted versus 4\emph{f} spin, $S$,
of the studied lanthanide metals. (b) Effective exchange-coupling
constant, $J_{\mathrm{eff}}$, and lattice constant, $a$, plotted
versus 4\emph{f} occupancy, $n_{4f}$, of various lanthanide
meals.}
\end{center}
\end{figure}

Table~\ref{tab1} summarizes the fit results for the energies,
$E_{\uparrow}$ and $E_{\downarrow}$, and the widths,
$\Gamma_{\uparrow}$ and $\Gamma_{\downarrow}$, of the majority
($\uparrow$) and minority ($\downarrow$) surface states as
obtained from numerous STS spectra for each lanthanide metal. The
exchange splittings, $\Delta_{ex} = E_{\downarrow} -
E_{\uparrow}$, are plotted in Fig.~\ref{fig2}(a) versus the
4\emph{f}-spin quantum number $S$. A systematic deviation from a
linear dependence is clearly observed. Furthermore, Nd exhibits a
much smaller exchange splitting than Er, although both elements
have the same 4\emph{f} spin.

\begin{table*}
\begin{center}
\begin{tabular}{ccccccccccccccccc} \hline \hline
 & La & \multicolumn{2}{c}{Nd} & \multicolumn{2}{c}{Gd} & \multicolumn{2}{c}{Tb} & \multicolumn{2}{c}{Dy} & \multicolumn{2}{c}{Ho} & \multicolumn{2}{c}{Er} & \multicolumn{2}{c}{Tm} & Lu \\

 & & $\uparrow$ & $\downarrow$ &  $\uparrow$ & $\downarrow$ &  $\uparrow$ & $\downarrow$ &  $\uparrow$ & $\downarrow$ &  $\uparrow$ & $\downarrow$ &
$\uparrow$ & $\downarrow$ &  $\uparrow$ & $\downarrow$  & \\
\hline
$\Gamma$ (meV) \hspace{1mm} & 49 \hspace{1mm} & 70 & 120 \hspace{1mm} & 33 & 134 \hspace{1mm} & 22 & 184 \hspace{1mm} & 19 & 170 \hspace{1mm} & 19 & 125 \hspace{1mm} & 30 & 146 \hspace{1mm} & 29 & 190 \hspace{1mm} & 5.0 \\
$E - E_{F}$ (meV) \hspace{1mm} & 130 \hspace{1mm} & -18 & 187 \hspace{1mm} & -191 & 470 \hspace{1mm} & -111 & 474 \hspace{1mm} & -102 & 402 \hspace{1mm} & -90 & 347 \hspace{1mm} & -51 & 318 \hspace{1mm} & -19 & 240 \hspace{1mm} & 2.0 \\
\hline \hline
\end{tabular}
\caption{\label{tab1} Lifetime width, $\Gamma$, and energy, $E -
E_{F}$, of exchange-split (0001)-surface states on trivalent
lanthanide metals. The estimated errors of the energies are $\pm
10\, \mbox{meV}$, in case of Lu $\pm 1\, \mbox{meV}$. The errors
of the $\Gamma$ values are plotted in Fig.~\ref{fig3}. Majority
and minority states are denoted by $\uparrow$ and $\downarrow$,
respectively.}
\end{center}
\end{table*}

We introduce an effective exchange-coupling constant,
$J_{\mathrm{eff}} = \Delta_{ex}/S$, and plot the resulting
$J_{\mathrm{eff}}$ in Fig.~\ref{fig2}(b) versus $n_{4f}$. In
contrast to previous results \cite{Wes01,Bau02},
$J_{\mathrm{eff}}$ is not found to be constant for all lanthanide
metals but increases almost linearly with increasing $n_{4f}$.
Another quantity that changes almost linearly with $n_{4f}$ as a
consequence of lanthanide contraction is the lattice constant $a$
[see open symbols in Fig.~\ref{fig2}(b)]. This suggests a relation
between $J_{\mathrm{eff}}$ and $a$ of the form $\Delta
J_{\mathrm{eff}} \propto - \Delta a$, which is plausible, since
the exchange interaction of the outer (5\emph{d}6\emph{s})
electronic states with the 4\emph{f} core states will depend on
the overlap of these states. Along this line, the smaller exchange
splitting observed for Nd as compared to Tm can be understood,
fitting nicely into the general trend. It should be noted,
however, that Nd has a rather complex spin structure at 10 K, with
a vanishing net moment in the (0001) planes \cite{For92}, whereas
the other heavy lanthanide metals exhibit ferromagnetic coupling
of the 4\emph{f} spins within the (0001) planes. This can also
contribute to the smaller exchange splitting observed for
Nd(0001), since the surface state, although laterally highly
localized, overlaps with neighboring atoms and, in this way,
interacts with a reduced average 4\emph{f} spin in the Nd case.

\begin{figure}
\begin{center}
\includegraphics{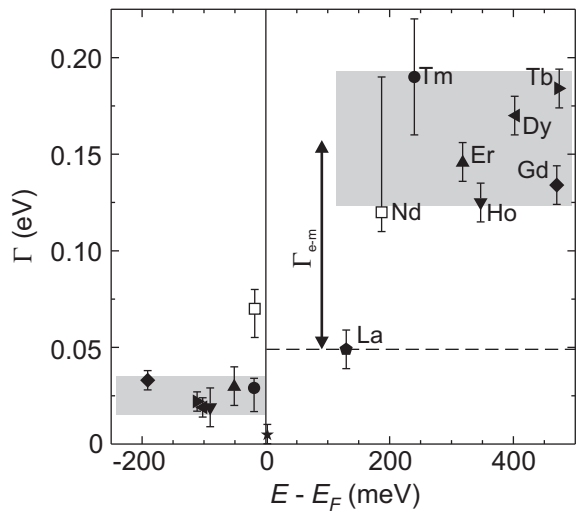}
\caption{\label{fig3} Lifetime width, $\Gamma$, versus
surface-state energies, $E - E_{F}$, for the trivalent lanthanide
metals according to measured values listed in Table \ref{tab1}.}
\end{center}
\end{figure}

In the following, we discuss the lifetime widths of surface
states. In Fig.~\ref{fig3}, the $\Gamma$ values, listed in Table
\ref{tab1}, are plotted versus surface-state energies. As
indicated by the two shaded areas, the lifetime widths of the
majority states below $E_{F}$ are considerably smaller than those
of the minority states above $E_{F}$. In a previous study of Gd
and Ho \cite{Bau02,Reh03}, the enhanced lifetime widths were
attributed to the larger absolute energies, $|E-E_{F}|$, of the
unoccupied minority states, which, according to 3-dimensional
Fermi-liquid theory, would result in increased e-e scattering
rates: $\Gamma_{e-e} = 2 \beta \cdot (E - E_{F})^2$ \cite{Qui58}.
In this way, $2 \beta$ values between $0.2$ and $0.5\,
\mbox{eV}^{-1}$ were obtained, which are, however, up to an order
of magnitude larger than the $2 \beta$ values reported for other
materials \cite{Ech04}, including Yb, a divalent lanthanide metal
with a $4f^{14}$ configuration. In case of Yb, $2 \beta \approx
0.08\, \mbox{eV}^{-1}$ was recently determined from lifetime
widths of quantum-well states in thin Yb(0001) films \cite{Weg05}.
Although we cannot completely rule out the possibility of large $2
\beta$ values, the present results, based on a systematic study of
trivalent lanthanide metals, renders the previous conclusion
questionable.

For surface states that lie in the center of a bandgap of the
projected bandstructure, interband-transition rates are expected
to be small since such transitions involve large momentum
transfers, $\Delta k$, and the Coulomb interaction scales with
$\Delta k^{-2}$ \cite{Ech04}. Intraband transitions, on the other
hand, may give significant, in some cases even dominant
contributions to lifetime widths \cite{Kli00}. In the tunneling
process, holes ($U < 0$) and electrons ($U > 0$) are predominantly
excited at $\overline{\Gamma}$, i.e.\ at the maximum of the
surface-state band. While an excited electron in an unoccupied
surface-state band can decay into lower-lying electronic states of
the same band, there are no intraband decay channels available for
a hole in an occupied surface-state band at $\overline{\Gamma}$.
We therefore conclude that the dominant contribution to lifetime
widths of the occupied surface states is e-ph scattering. For Lu,
the surface-state energy is smaller than the Debye energy so that
even e-ph scattering is suppressed \cite{Bau02}. From
temperature-dependent measurements of Gd surface states, the
zero-temperature contribution of e-ph scattering to the lifetime
width was determined as $\Gamma_{e-ph} = (22 \pm 5)\, \mbox{meV}$
\cite{Reh03}. Thus, e-ph scattering accounts for approximately
70\% of the total lifetime width. Similar values can be expected
for the other lanthanide metals.

For unoccupied surface states, intraband e-e scattering will be
enhanced, explaining the increased lifetime width of the
unoccupied La-surface state by about 25 meV as compared to those
of the occupied surface states.

So far, we have not taken into account e-m scattering of surface
states due to exchange interaction with the 4\emph{f} spins. For
the occupied surface states, the same argument as given above for
e-e scattering applies, i.e.\ the number of states into which an
excited hole at or close to the surface-state band maximum can
scatter is limited. Furthermore, due to the high degree of spin
polarization of the surface state bands, intraband spin-flip
scattering is suppressed, even for excited states below the
surface-state band maximum. Previous studies of Gd based on
spin-resolved PE found an e-m-scattering contribution of $\approx
14\, meV$ to the lifetime width of the majority-spin state, but
only if the state is more than 25 meV -- the highest magnon energy
-- below the band maximum \cite{All01,Fed02}. Due to the small
dispersions of the surface-state bands, most excited holes will be
closer to the band maximum so that the actual contribution of e-m
scattering will be smaller \cite{Bau02}.

For unoccupied surface states on magnetic lanthanide metals, both
intraband and interband spin-flip transition rates will be
enhanced. This explains the large lifetime widths compared to
those of the occupied surface states and also to that of the
unoccupied surface state on non-magnetic La metal. Again,
intraband spin-flip transitions are only possible if the spin
polarization of the surface-state band is less than 100\%, e.g.,
due to spin-orbit coupling \cite{All01,Fed02}. As for interband
transitions, it is interesting to note that for Gd, and presumably
also for the other magnetic lanthanide metals, the unoccupied
minority surface-state band lies outside the bandgap of the
projected majority bandstructure, whereas the occupied minority
surface states are located within the total, spin-averaged bandgap
\cite{Kur02}. Therefore, spin-flip transitions by magnon emission
do not require large momentum transfers, as spin-conserved
transitions do, which could explain the enhanced scattering rates.
If we make the assumption that the differences in lifetime widths
of the unoccupied surface states on magnetic lanthanide metals, on
the one hand, and the lifetime width of the surface state on
La(0001), on the other hand, is solely due to e-m scattering, the
e-m contribution amounts to $\approx (100 \pm 25)\, \mbox{meV}$;
the error bar reflects the scattering of data points for different
metals (see Fig.~\ref{fig3}).

Finally, we like to mention that the rather large lifetime width
of the occupied surface state on Nd is probably also due to e-m
scattering. While the exchange-split surface states of the heavy
lanthanide metals are nearly 100\% spin polarized, the net
polarization of Nd surface states vanishes due to the
antiferromagnetic spin order within the (0001)-surface plane,
allowing for enhanced spin-flip intraband scattering.

In summary, the present work provides detailed information on the
electronic structure of surface states on lanthanide metals, in
particular on their energies, exchange splitting, and dynamics. At
this point, theoretical calculations are needed to corroborate the
experimental findings and conclusions regarding the influence of
lanthanide contraction on exchange splitting as well as the
contribution of e-ph, inter- and intraband e-e, and e-m scattering
to the lifetime widths of the surface states.

%
%
This work was supported by the  Deutsche Forschungsgemeinschaft,
project Sfb-290/TPA6. A.B.\ acknowledges support within the
Heisenberg program of the Deutsche Forschungsgemeinschaft.

\bibliography{lifetime_1v}




\end{document}